\documentclass[twocolumn,showpacs,preprintnumbers,amsmath,amssymb,prl,superscriptaddress]{revtex4}
\usepackage{times}
\usepackage{amsfonts}
\usepackage{mathrsfs}
\usepackage{graphicx}
\usepackage{dcolumn}
\usepackage{bm}
\usepackage{color}

\usepackage[colorlinks,bookmarks=false,citecolor=blue,linkcolor=red,urlcolor=blue]{hyperref}
\usepackage{multirow}
\usepackage{amsmath}

\usepackage{bbm}
\usepackage{pifont}
\usepackage[normalem]{ulem}

\newcommand{\RNum}[1]{\uppercase\expandafter{\romannumeral #1\relax}}

\newcommand{\Rom}[1]{ \uppercase\expandafter{\romannumeral#1}}

\newcommand{\kk}{{\bm{\mathrm{k}}}}

\newcommand{\rr}{{\bf{r}}}

\newcommand{\bmS}{{\bm{\mathrm{S}}}}
\newcommand{\bms}{{\bf{s}}}

\newcommand{\RR}{{\bf{R}}}

\newcommand{\ttau}{\bm{\tau}}

\newcommand{\bma}{\bm{a}}

\newcommand{\ee}{\bm{e}}

\definecolor{darkblue}{rgb}{0.15,0.25,0.6}
\definecolor{ZYcolor}{rgb}{0.1,0.5,0.4}
\definecolor{darkred}{rgb}{0.65,0.2,0.15}

\def\be{\begin{equation}}       \def\ee{\end{equation}}
\def\bea{\begin{eqnarray}}      \def\eea{\end{eqnarray}}

\begin{document}

\title{Diversified Ruderman-Kittel-Kasuya-Yosida Interactions in a Nonsymmorphic Crystal}

\author{Zhongyi Zhang}
\affiliation{Beijing National Laboratory for Condensed Matter Physics and Institute of Physics, Chinese Academy of Sciences, Beijing 100190, China}
\affiliation{University of Chinese Academy of Sciences, Beijing 100049, China}

\author{Shengshan Qin}\email{qinshengshan@ucas.ac.cn}
\affiliation{Kavli Institute for Theoretical Sciences and CAS Center for Excellence in Topological Quantum Computation, University of Chinese Academy of Sciences, Beijing 100190, China}
\affiliation{University of Chinese Academy of Sciences, Beijing 100049, China}

\author{Chen Fang}
\affiliation{Beijing National Laboratory for Condensed Matter Physics and Institute of Physics, Chinese Academy of Sciences, Beijing 100190, China}
\affiliation{Kavli Institute for Theoretical Sciences and CAS Center for Excellence in Topological Quantum Computation, University of Chinese Academy of Sciences, Beijing 100190, China}

\author{Jiangping Hu}
\affiliation{Beijing National Research Center for Condensed Matter Physics,
and Institute of Physics, Chinese Academy of Sciences, Beijing 100190, China}
\affiliation{Kavli Institute for Theoretical Sciences and CAS Center for Excellence in Topological Quantum Computation, University of Chinese Academy of Sciences, Beijing 100190, China}
\affiliation{South Bay Interdisciplinary Science Center, Dongguan, Guangdong Province, China}

\author{Fu-chun Zhang}
\affiliation{Kavli Institute for Theoretical Sciences and CAS Center for Excellence in Topological Quantum Computation, University of Chinese Academy of Sciences, Beijing 100190, China}
\affiliation{Collaborative Innovation Center of Advanced Microstructures, Nanjing University, Nanjing 210093, China}

\date{\today}

\begin{abstract}
We show that there are diversified Ruderman-Kittel-Kasuya-Yosida (RKKY) interactions between magnetic impurities, mediated by itinerant electrons, in a centrosymmetric crystal respecting a nonsymmorphic space group. We take the $P4/nmm$ space group as an example. We demonstrate that the different type of interactions, including the Heisenberg-type, the Dzyaloshinskii-Moriya (DM)-type, the Ising-type and the anisotropic interactions, can appear in accordance with the positions of the impurities in the real space. Their strengths strongly depend on the location of the itinerant electrons in the reciprocal space.
The diversity stems from the position-dependent site groups and the momentum-dependent electronic structures guaranteed by the nonsymmorphic symmetries. Our study unveils the role of the nonsymmorphic symmetries in affecting magnetism, and suggests that the nonsymmorphic crystals can be promising platforms to design magnetic interactions.
\end{abstract}

\maketitle
\textit{Introduction.}
In noncentrosymmetric systems, the Rashba spin-orbit coupling (SOC) arises due to the absence of the inversion symmetry \cite{rashba1960properties}.
The Rsshba SOC leads to the spin-momentum lock, which plays an important role in lots of exotic quantum phenomena \cite{armitage2018weyl,chiu2016classification,yan2017topological}, such as the spin-orbit torques \cite{garello2013symmetry,yu2014switching}, the singlet-triplet mixing superconductivity \cite{yip2014noncentrosymmetric,smidman2017superconductivity}, the various topological phases of matter \cite{burkov2011weyl,hosur2013recent,PhysRevLett.100.096407,das2012zero,sau2010generic,PhysRevLett.105.077001,PhysRevLett.105.177002,PhysRevB.84.201105}, etc.
Especially, when the Rashba SOC is encoded with magnetism, the DM magnetic interactions can be induced \cite{bruno1995theory,yafet1987ruderman,dzyaloshinsky1958thermodynamic,moriya1960anisotropic,moriya1960new}.
The DM interaction has been extensively studied in recent years, because of its essential role in inducing the magnetic skyrmion \cite{wang2017rkky,chang2015rkky,valizadeh2016anisotropic,zhu2011electrically,everschor2018perspective,dai2013skyrmion,back20202020,kang2015transport,muhlbauer2009skyrmion}, the topological magnon \cite{li2016weyl,fransson2016magnon,owerre2016first,owerre2017magnonic} and the spin helix \cite{bernevig2006exact,koralek2009emergence}.
These interesting properties make the materials with large DM interaction be promising candidates for next-generation spintronics \cite{fert2017magnetic,kang2016skyrmion}.

Recently, a class of centrosymmetric systems, which have the so-called local inversion-symmetry-breaking effect \cite{zhang2014hidden,yuan2019uncovering,qin2022topological,wu2017direct,lin2021skyrmion,hayami2022skyrmion,agterberg2017resilient}, have attracted great research interest.
In such systems, the inversion center is off the lattice sites, making the Rashba SOC allowed even though the system being globally inversion symmetric.
Interestingly, in such systems there exists the net spin-momentum lock for electrons from certain subsystems, but the effect compensates for electrons from different subsystems.
Correspondingly, the DM magnetic interaction can be expected in such systems.
Moreover, besides magnetism it has been demonstrated that the local inversion-symmetry-breaking effect may be essential in the odd-parity superconductivity \cite{PhysRevB.105.L020505,khim2021field,qin2022spin} and the topological superconductivity \cite{fischer2022superconductivity,qin2022topological}.

The RKKY interaction, which plays a central role in the diluted magnetic semiconductors, is an indirect exchange interaction between magnetic impurities mediated by itinerant electrons.
It provides an another scheme for the DM magnetic interaction \cite{dugaev2006exchange,hosseini2015ruderman,liu2009magnetic,sun2017rkky,wang2022rkky}.
In this work, we present a detailed theoretical investigation on the RKKY interaction mediated by itinerant electrons in a centrosymmetric crystal respecting the nonsymmorphic space group $P4/nmm$. We demonstrate that though the itinerant electrons respect the inversion symmetry in the system, RKKY interactions including the Heisenberg, the DM, the Ising and the anisotropic terms, can be induced. Moreover, the specific forms of the RKKY interaction varies in accordance with the positions of the impurities in the real space, and the strength of the interaction varies according to the locations of the itinerant electrons in the reciprocal space. We discuss the rich and electronically controllable spin configurations in such systems.

\begin{figure}[!htbp]
  \centering
  \includegraphics[width=1\linewidth]{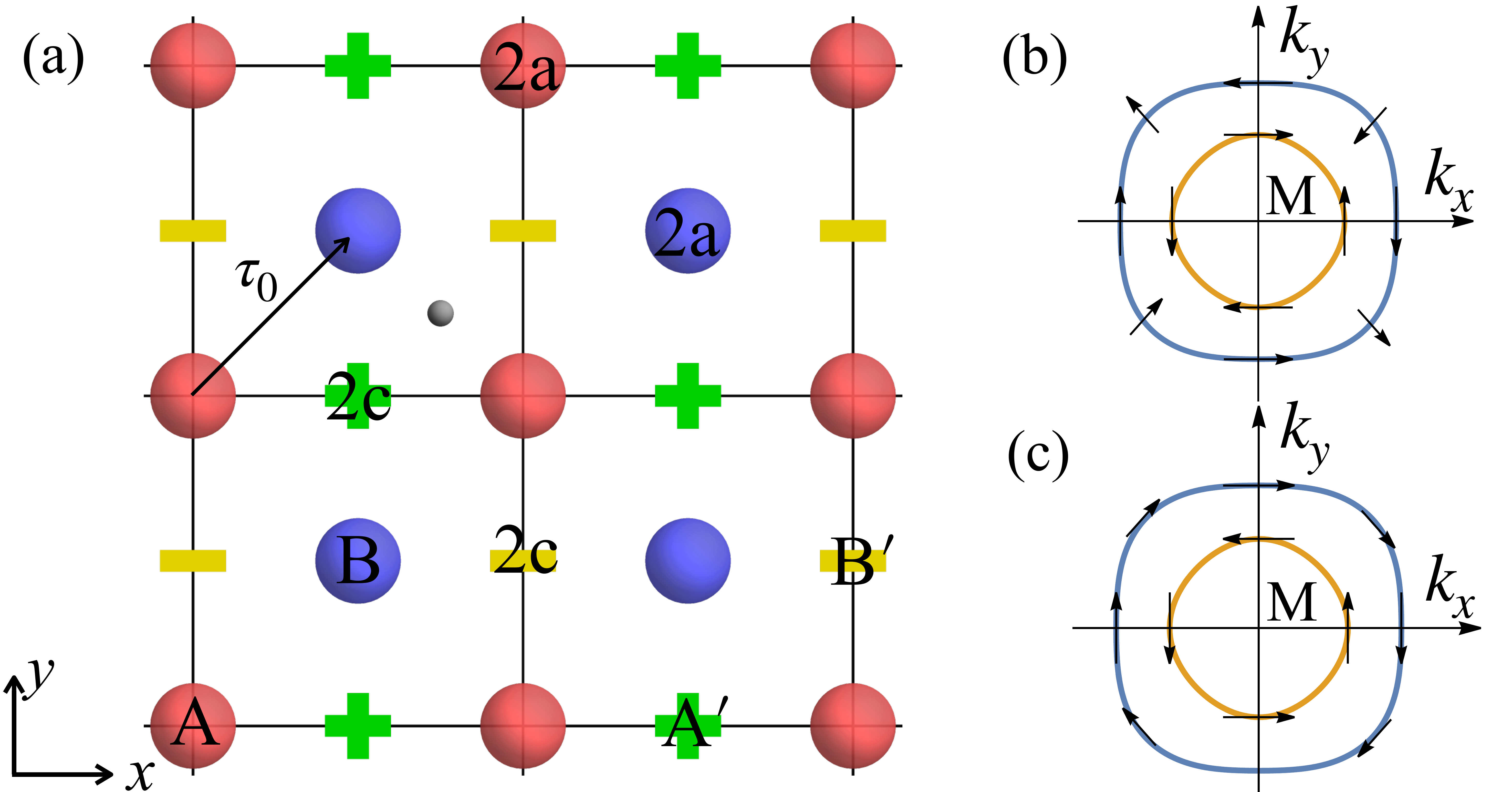}
  \caption{\label{fig1}(color online) (a) A quasi-2D lattice respects the space group $P4/nmm$. The red and blue circles labeled by $A$ and $B$ (green plus and yellow minus labeled by $A^\prime$ and $B^\prime$) represent the two $2a$ ($2c$) Wyckoff positions related by the nonsymmorphic symmetries, at which the $D_{2d}$ ($C_{4v}$) point group preserves. The gray circle is the inversion center at the middle between the two nearest-neighbour $2a$ ($2c$) Wyckoff positions.
  (b) and (c) sketch the spin polarizations, $\langle\bms_{A(A^\prime)}\rangle-\langle\bms_{B(B^\prime)}\rangle)$, on the Fermi surfaces near M in the BZ, corresponding to the Hamiltonian in Eq.~\eqref{2a eff H} and Eq.~\eqref{2c eff H} respectively. Here, $\langle\bms_{A(A^\prime)}\rangle$ and $\langle\bms_{B(B^\prime)}\rangle$ are the spin polarizations contributed by the orbitals at the two different $2a$ ($2c$) Wyckoff positions. Notice that the inversion symmetry and the time-reversal symmetry demand $\langle\bms_{A(A^\prime)}\rangle = -\langle\bms_{B(B^\prime)}\rangle)$ at each $\kk$ point. }
\end{figure}

\textit{Space group $P4/nmm$.}
We first briefly review the nonsymmorphic space group $\mathcal{G} = P4/nmm$. The group $\mathcal{G}$ has 16 symmetry operations in its quotient group $\mathcal{G}/T$, satisfying a special structure \cite{PhysRevX.3.031004}
\begin{equation}\label{little group at M}
\mathcal{G}/T  \cong D_{2d}\otimes Z_2 \cong C_{4v}\otimes Z_2,
\end{equation}
where $D_{2d}$, $C_{4v}$ and $Z_2$ are three point groups defined at different positions. To have a more intuitive impression, we consider a quasi-2D lattice shown in Fig.~\ref{fig1}(a). In the lattice, the Wyckoff positions $2a$ and $2c$ are the fixed points preserving the point group $D_{2d}$ and $C_{4v}$ respectively. The $Z_2$ group in Eq.~\eqref{little group at M} is a two-element group including the inversion symmetry which switches the two $2a$ ($2c$) Wyckoff positions in Fig.~\ref{fig1}(a). According to Eq.~\eqref{little group at M}, it is obvious that group $\mathcal{G}/T$ can be generated by the generators of $D_{2d}$ ($C_{4v}$) and $Z_2$, which can be chosen as  $\{ M_y | {\bf 0} \}$, $\{ S_{4z} | {\bf 0} \}$ ($\{ M_{xy} | \ttau_0 \}$, $\{ C_{4z} | \ttau_0 \}$), and $\{ I | {\bf \tau_0} \}$ respectively \cite{footnote_symm}.

\textit{Itinerant electrons in group $P4/nmm$.}
For a nonsymmorphic crystal, its group structure, i.e. the commutation relations between the symmetry operations, varies with the momentum in the reciprocal space \cite{PhysRevB.91.161105,zak1960method,PhysRevB.91.155120}. Actually, at the BZ center, i.e. the $\Gamma$ point $(0, 0)$, the system in Fig.~\ref{fig1}(a) respects the $D_{4h}$ point group \cite{footnote1}; while it respects a little group which is not isomorphic to any point group at the BZ corner, i.e. the M point $(\pi, \pi)$ \cite{cvetkovic2013space}.
The momentum-dependent group structures indicate the momentum-dependent properties of the itinerant electrons in a nonsymmorphic group.

To describe the itinerant electrons in group $P4/nmm$ specifically, we derive the low-energy effective theory. We start with the BZ corner. A standard group theory analysis shows that, the group $P4/nmm$ merely has one single 4D irreducible spinful representation at the M point, meaning that all the energy bands in the system are fourfold degenerate and respect the same effective theory at M in the spinful condition \cite{qin2022symmetry,smidman2017superconductivity}. Considering the constraints of the crystalline symmetries and the time-reversal symmetry, we obtain the effective theory as follows \cite{qin2022spin,qin2022symmetry,smidman2017superconductivity}
\begin{equation}\label{2a eff H}
\mathcal{H}_{\mathrm{M,2a}}(\kk)=m(\kk)s_0\sigma_0+\lambda k_x s_2\sigma_3+\lambda k_y s_1\sigma_3+t^\prime k_xk_y s_0\sigma_1,
\end{equation}
where $m(\kk)=t(k_x^2+k_y^2)$, and $t,t^\prime,\lambda$ are all constants. In Eq.~\eqref{2a eff H}, $s_i$ and $\sigma_i$ $(i=0,1,2,3)$ are Pauli matrices in the spin and sublattice spaces respectively. To have an intuitive impression on the effective theory, one can assume a single $s$ orbital at each $2a$ Wyckoff position in Fig.~\ref{fig1}(a). $t$ $(t^\prime)$ can be understood as the hopping between intrasublattice (intersublattice) nearest-neighbour sites, and $\lambda$ represents the Rashba SOC stemming from the mismatch between the lattice sites and the inversion center. According to Eq.~\eqref{little group at M}, one can also set the orbital at the $2c$ Wyckoff positions in Fig.~\ref{fig1}(a), in which condition the effective theory at M reads as (details in SM \cite{SM})
\begin{equation}\label{2c eff H}
\mathcal{H}_{\mathrm{M,2c}}(\kk)=m(\kk)s_0\sigma_0+\lambda k_x s_2\sigma_3-\lambda k_y s_1\sigma_3+t^\prime k_xk_y s_0\sigma_1,
\end{equation}
where the coefficients can be understood in a similar way with these in $\mathcal{H}_{\mathrm{M,2a}}$. Notice that $\mathcal{H}_{\mathrm{M,2a}}$ and $\mathcal{H}_{\mathrm{M,2c}}$ describes the same fourfold band degeneracy at M. However, corresponding to the different positions where the orbitals locate, the physical observables are different. For example, we calculate the spin textures on the Fermi surfaces for the effective theories in Eq.~\eqref{2a eff H} and Eq.~\eqref{2c eff H} and show the results in Fig.~\ref{fig1}(b) and Fig.~\ref{fig1}(c) respectively. As shown, the spin texture in Fig.~\ref{fig1}(b) (Fig.~\ref{fig1}(c)) is $D_{2d}$ ($C_{4v}$) symmetric, in accordance with the site group at the $2a$ ($2c$) Wyckoff positions chosen for $\mathcal{H}_{\mathrm{M,2a}}$ ($\mathcal{H}_{\mathrm{M,2c}}$). The different theories in Eqs.~\eqref{2a eff H}\eqref{2c eff H} demonstrate the profound roles of the real-space positions on the RKKY interactions.

At the $\Gamma$ point, the system respects the point group $D_{4h}$ and the corresponding effective theory takes the form (details in SM \cite{SM})
\begin{equation}\label{2a eff G}
\mathcal{H}_{\mathrm{\Gamma},2a}(\kk)=m(\kk)s_0\sigma_0+\lambda k_x s_2\sigma_3+\lambda k_y s_1\sigma_3+t^\prime  s_0\sigma_1,
\end{equation}
where the basis is the same with that for the effective theory $\mathcal{H}_{\mathrm{M,2a}}$ in Eq.~\eqref{2a eff H}. For the basis in Eq.~\eqref{2c eff H}, the effective theory at $\Gamma$ becomes
\begin{equation}\label{2c eff G}
\mathcal{H}_{\mathrm{\Gamma,2c}}(\kk)=m(\kk)s_0\sigma_0+\lambda k_x s_2\sigma_3-\lambda k_y s_1\sigma_3+t^\prime s_0\sigma_1,
\end{equation}
Apparently, $\mathcal{H}_{\mathrm{\Gamma},2a(2c)}$ describes two Kramers' doublets which is different from the condition at M.

Comparing the effective theories near $\Gamma$ and M, it can be found that, near $\Gamma$ the effective SOC on the energy bands is vanishing small because of the finite $t^\prime  s_0\sigma_1$ term, while the system is nearly a direct product of two Rashba electron gas systems due to the dominating SOC term near M. Such difference implies the different RKKY interactions corresponding to itinerant electrons at the BZ center and corner.

\textit{RKKY interaction in $P4/nmm$.}
We consider two magnetic impurities $\bmS_i$ ($i = 1, 2$) located at $\RR_i$ in the system in Fig.~\ref{fig1}(a). The magnetic impurities interact with the itinerant electrons through the $s-d$ interaction \cite{kasuya1956theory}
\begin{equation}\label{sd}
\mathcal{H}_{sd} = -J [ ( \bmS_1 \cdot \bms ) \sigma_\alpha \delta(\rr-\RR_1) + ( \bmS_2 \cdot \bms ) \sigma_\beta \delta(\rr-\RR_2) ],
\end{equation}
where $J$ is the strength of the exchange coupling.
In Eq.~\eqref{sd}, $\sigma_{\alpha/\beta}$ denotes the sublattice locked to the position $\RR_i$ which takes the form $(\sigma_1 + \sigma_3)/2$ or $(\sigma_1 - \sigma_3)/2$. The RKKY interaction between the impurities can be obtained by integrating out the itinerant electrons
\begin{equation}\label{RKKY}
\begin{aligned}
H_{\mathrm{RKKY}}=&-\frac{J^2}{\pi}\mathrm{Im}\int_{\omega <\mu} \mathrm{d}\omega\ \mathrm{Tr} [(\bmS_2\cdot\bms)\sigma_\beta G(\RR_2,\RR_1,\omega )\\
&\times(\bmS_1\cdot\bms)\sigma_\alpha G(\RR_1,\RR_2,\omega )],
\end{aligned}
\end{equation}
where $\mu$ is the Fermi energy, $\mathrm{Tr}$ represents the trace over the degrees of the itinerant electrons, and $G(\RR_i,\RR_j,\omega )$ is the real-space Green function for the itinerant electrons with $\omega$ the frequency. After some algebra, we find the RKKY interaction in Eq.~\eqref{RKKY} takes the form
\begin{equation}\label{RKKY express}
H_{\mathrm{RKKY}}(\RR,\mu)=\Lambda \bmS_1\cdot\bmS_2 +   {\bf D}\cdot (\bmS_1\times\bmS_2) +\sum_{i,j}T_{ij}S_{1i}S_{2j},
\end{equation}
with $\RR = \RR_1 - \RR_2$, ${\bf D}=(\mathrm{D}_1,\mathrm{D}_2)$ and $i,j=x,y,z$.
In Eq.~\eqref{RKKY express}, the first term is the Heisenberg-type, the second term is the DM-type, the last term includes the Ising-type terms ($i = j$) and the anisotropic terms ($i \neq j$). Its specific form and strength depend on the location of the impurities in the real space and the location of the itinerant electrons in the reciprocal space.

\textit{Symmetry constraints.}
Before going to the details, we first consider the symmetry constraints on the RKKY interaction. As we shall show, the Heisenberg and DM terms in Eq.~\eqref{RKKY express} dominate other terms. Therefore, we focus on these two terms in the analysis. The Heisenberg-type magnetic interaction, $\bmS_1\cdot\bmS_2$, is always invariant under the crystalline symmetries. The DM-type interaction, ${\bf D}\cdot (\bmS_1\times\bmS_2)$, is characterized by the so-called DM vector, ${\bf D}$. Moreover, to guarantee the term a scalar, ${\bf D}$ must be a pseudovector. A pseudovector behaves like a vector under the proper spatial symmetry whereas remain unchanged under inversion symmetry. This imposes strict constraints on the DM interaction \cite{moriya1960new} and makes it strongly depend on the positions of the impurities in Fig.~\ref{fig1}(a).

\begin{figure}[!htbp]
  \centering
  \includegraphics[width=1\linewidth]{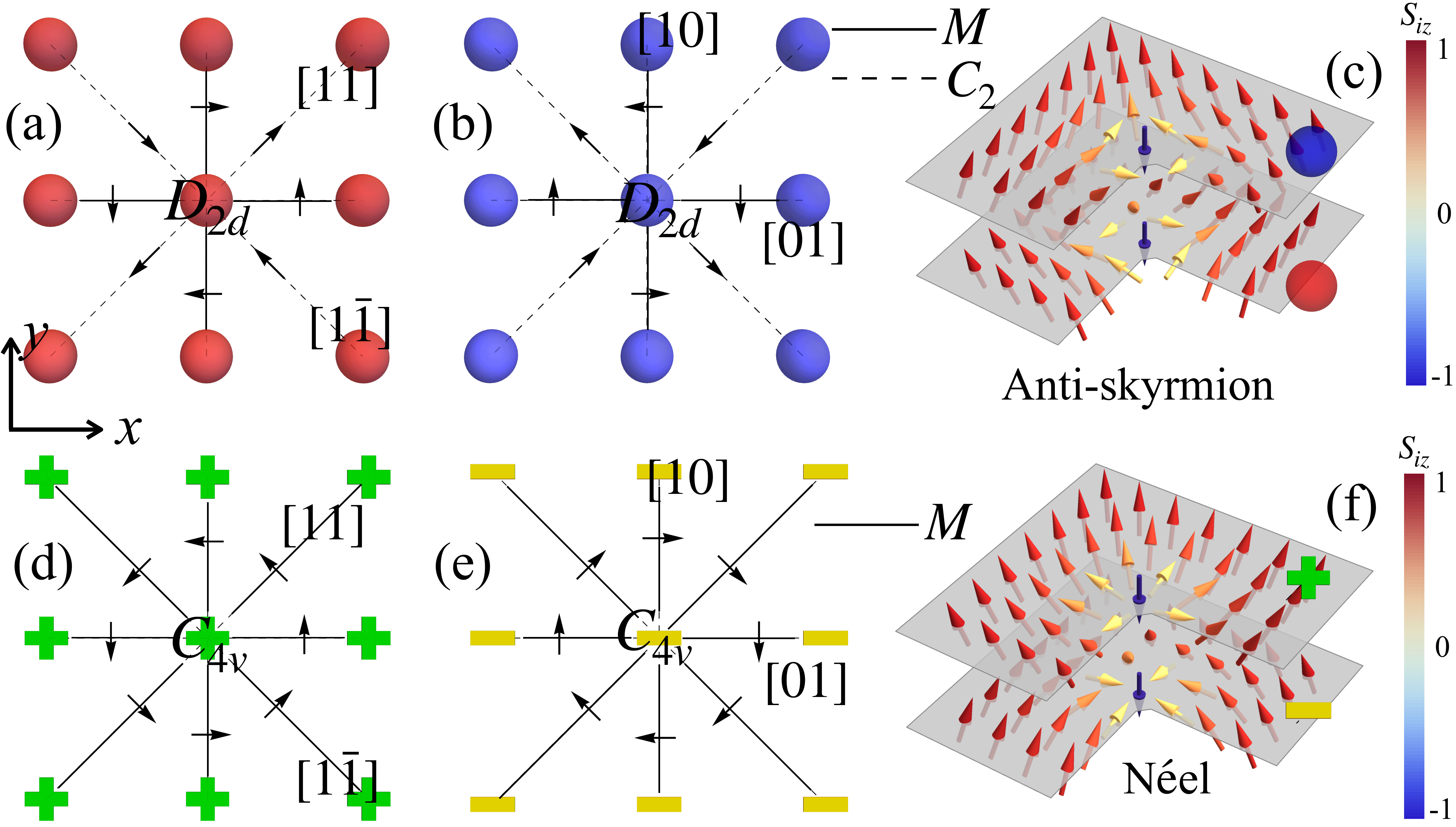}
  \caption{\label{fig2}(color online) (a)(b) and (d)(e) sketch the directions of the DM vectors ${\bf D}$ in Eq.~\eqref{RKKY express}, corresponding to the condition that the two impurities locate in the same sublattice at the $2a$ and $2c$ Wyckoff positions in Fig.~\ref{fig1}(a) respectively. (a) ((d)) and (b) ((e)) represent the $A$ ($A^\prime$) and $B$ ($B^\prime$) sublattices related by the inversion symmetry in Fig.~\ref{fig1}(a).
  The DM interactions in (a)(b) and (d)(e) support the antiskyrmion-type and N$\acute{{\rm e}}$el-type spin textures shown in (c) and (f) respectively.
  }
\end{figure}

We first consider the conditions where the two impurities locate in the same sublattice at the $2a$ Wyckoff positions, i.e. the $A$ ($B$) sublattice in Fig.~\ref{fig1}(a). As such sites are $D_{2d}$ invariant, the DM vectors must be compatible with the $D_{2d}$ group. For instance, the mirror symmetry $\{{M}_x|{\bf0}\}$ demands ${\bf D}$ perpendicular to $\RR_1-\RR_2$, if the displacement of the two impurities is along the $[10]$ direction; and ${\bf D}$ is parallel to $\RR_1-\RR_2$ in the $[11]$ direction, due to the $C_2$ rotation along $[11]$. Considering these constraints, we sketch the DM vectors in Fig.~\ref{fig2}(a) for the condition where the two impurities locate in the A sublattice at the $2a$ Wyckoff positions, and the DM vectors in the B sublattice is shown in Fig.~\ref{fig2}(b) as the two sublattices are related by the inversion symmetry. If the two impurities locate in the same sublattice at the $2c$ Wyckoff positions, i.e. the $A^\prime$ ($B^\prime$) sublattice in Fig.~\ref{fig1}(a), the DM vectors must be compatible with the $C_{4v}$ group and we sketch the results in Figs.~\ref{fig2}(d)(e).

It is worth pointing out that, in the above conditions the DM interactions are allowed because the inversion symmetry is absence within the same sublattice at $2a$ ($2c$) Wyckoff positions. When the two impurities locate in the different sublattices at the $2a$ ($2c$) Wyckoff positions, the inversion symmetry always exists and the DM interaction must vanish. The symmetry constraints on the DM interaction corresponding to other impurity configurations can be analyzed similarly. We note that, the above symmetry analysis does not depend on the location of the itinerant electrons in the BZ.

\begin{figure}[b]
  \centering
  \includegraphics[width=1.0\linewidth]{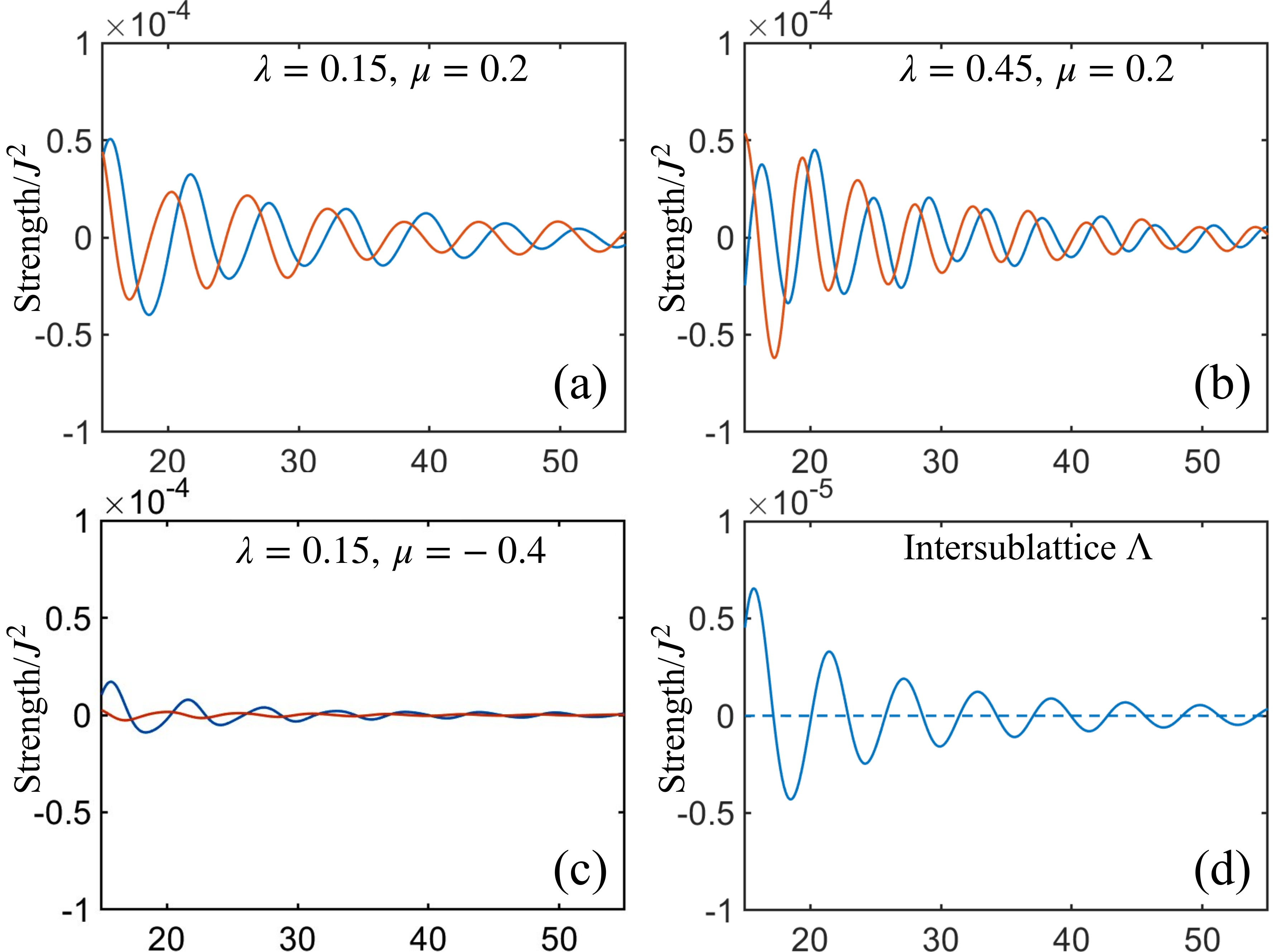}
  \caption{\label{fig3}(color online) The range function of the Heisenberg term $\Lambda$ (bule) and the second component of DM term $\mathrm{D}_2$ (orange) in Eq.~\eqref{RKKY express} in the condition where the two impurities locate at the $2a$ Wyckoff positions in Fig.~\ref{fig1}(a). In (a)(b)(c) the two impurities are in the same sublattice, satisfying $\RR \parallel [10]$. (a)(b) correspond to itinerant electrons near M, and (c) corresponds to the $\Gamma$ point. In (a) and (c), we tune the Fermi energy to make the itinerant electrons near M and $\Gamma$ have the same filling. In (d) the impurities are in different sublattices satifying $\RR \parallel [11]$, and we compare the Heisenberg terms mediated by itinerant electrons near the $\Gamma$ (dashed) and M (solid) points. In the calculations, the other parameters are set to be $t=1$, $t^\prime=0.7$.
  }
\end{figure}

\begin{figure}[t]
  \centering
  \includegraphics[width=1\linewidth]{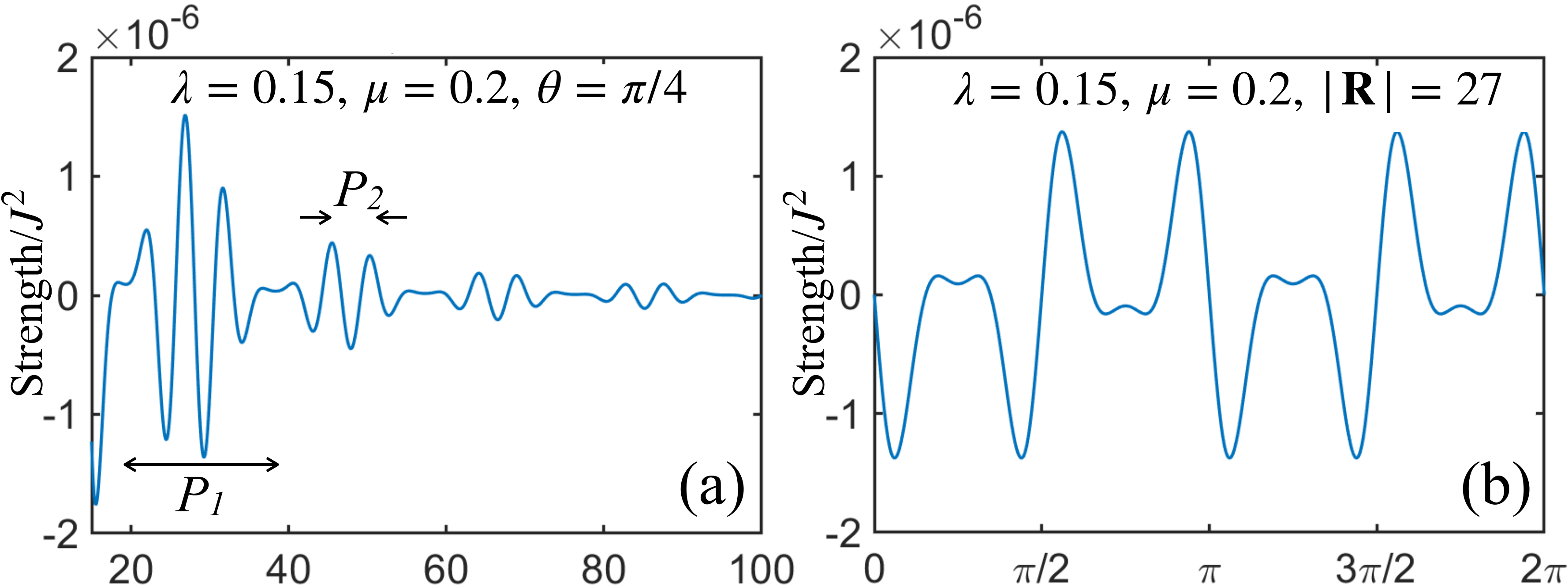}
  \caption{\label{fig4}(color online) The range function $T_{ij}$ for RKKY interaction in Eq.~\eqref{RKKY express} mediated by itinerant electrons near the M point, in the condition where the two impurities locate in the same sublattice at the $2a$ Wyckoff positions in Fig.~\ref{fig1}(a).
  (a) corresponds to the condition $\RR \parallel [11]$, where it satisfies $T_{xx} = T_{xy} = T_{yy}$.
  (b) shows $T_{xy}$ with respect to $\theta$ at $|\RR|=27$, with $\theta$ the polar angle defined according to the $x$ axis. In the calculations, the parameters are the same with these for Fig.~\ref{fig3}(a).
  }
\end{figure}

\textit{Numerical results.}
We simulate the range functions in Eq.~\eqref{RKKY express} numerically. Here, we only show the results corresponding to the condition where the impurities locate at the $2a$ Wyckoff positions in Fig.~\ref{fig1}(a). The results correspondingly to the $2c$ Wyckoff positions are similar and more details are presented in the SM \cite{SM}. Specifically, for impurities in the same sublattice, the range functions for the RKKY interactions in Eq.~\eqref{RKKY express} mediated by the itinerant electrons near the M and $\Gamma$ points take the form
\begin{equation}\label{range function1}
\begin{aligned}
\Lambda(\RR,\mu)&=-\frac{J^2}{\pi}\mathrm{Im}\int_{\omega <\mu} \mathrm{d}\omega\ (g_0^2+g_1^2+g_2^2),\\
{\mathrm{D}}_{1(2)}(\RR,\mu)&=-\frac{2 J^2}{\pi}\mathrm{Re}\int_{\omega <\mu} \mathrm{d}\omega\ g_0g_{2(1)},\\
T_{ij}(\RR,\mu)&=-\frac{J^2}{\pi}\mathrm{Im}\int_{\omega <\mu} \mathrm{d}\omega\ g_ig_j,\\
\end{aligned}
\end{equation}
where
\begin{equation}\label{range function2}
\begin{aligned}
&g_0 =\sum_{\alpha=\pm}\int \frac{\mathrm{d}^2\kk}{(2\pi)^2} e^{i\kk\cdot(\RR_1-\RR_2)}\frac{1}{2\Xi_{\theta}(\kk)}\frac{1}{\omega-E^{\theta}_{\alpha}(\kk)+i0^+},\\
&g_i =\sum_{\alpha=\pm}\int \frac{\mathrm{d}^2\kk}{(2\pi)^2} e^{i\kk\cdot(\RR_1-\RR_2)}\frac{1}{2\Xi_{\theta}(\kk)}\frac{\alpha\lambda k_i}{\omega-E^{\theta}_{\alpha}(\kk)+i0^+},
\end{aligned}
\end{equation}
with $i,j=1,2$, $k_{1(2)} = k_{x(y)}$, and $0^+$ being a positive infinitesimal. In Eq.~\eqref{range function2}, $E^{\theta}_{\pm}(\kk)= m(\kk)\pm\Xi_\theta(\kk)$ are the eigenenergies of the itinerant electrons near the $\theta = \mathrm{M}, \Gamma$ points, where $\Xi_\theta(\kk)=\sqrt{\lambda^2(k_x^2+k_y^2)+f_\theta^2}$ with $f_{\mathrm{M}} = t^\prime k_x k_y$ and $f_\Gamma = t^\prime$. If the two impurities are in the different sublattices, in Eq.~\eqref{RKKY express} only the Heisenberg term survives with the range function
\begin{equation}\label{range function3}
\Lambda(\RR,\mu) = -\frac{J^2}{\pi}\mathrm{Im}\int_{\omega <\mu} \mathrm{d}\omega\ g^{\prime2},\ \mathrm{D}_{i}(\RR,\mu)  = T_{ij}(\RR,\mu) = 0,
\end{equation}
where
\begin{equation}\label{range function4}
\begin{aligned}
&g^\prime =\sum_{\alpha=\pm}\int \frac{\mathrm{d}^2\kk}{(2\pi)^2} e^{i\kk\cdot(\RR_1-\RR_2+\ttau_0)}\frac{1}{2\Xi_{\theta}(\kk)}\frac{\alpha f_{\theta}}{\omega-E^{\theta}_{\alpha}(\kk)+i0^+}.
\end{aligned}
\end{equation}
Based on Eqs.~\eqref{range function1}$\sim$\eqref{range function4}, we plot the RKKY interactions in different cases in Fig.~\ref{fig3} and Fig.~\ref{fig4}.
Accordingly, one can find the following features of the RKKY interactions in the system.

The DM interaction can exist only when the two impurities are in the same sublattice, which is consistent with the symmetry analysis in the above. Moreover, the itinerant electrons near M are more in favor of the DM interaction which is comparable to the Heisenberg term, as indicated in Figs.~\ref{fig3}(a)$\sim$(c); and a larger $\lambda$ leads to the stronger DM term in comparison with the Heisenberg term, as shown in Figs.~\ref{fig3}(a)(b). The latter reflects the fact that, the local inversion symmetry breaking effect in the system is characterized by $\lambda$, i.e. the strength of the inversion symmetric Rashba SOC; and the former arises from the fact that, near $\Gamma$ the system is more like a conventional centrosymmetric system with vanishing samll effective Rashba SOC on the energy bands, while near M the system is nearly a direct product of two Rashba electron gas systems, as indicated in the effective theories in Eqs.~\eqref{2a eff H}\eqref{2a eff G}. Moreover, according to the effective Rashba SOC which can be featured by the spin polarizations on the energy bands shown in Fig.~\ref{fig1}(b), one can conclude that the smaller Fermi energy for itinerant electrons near M, i.e. the smaller $\kk$ near M, is better for the DM interaction.

In the condition where the two impurities locate in the different sublattices, only the Heisenberg interaction can exist and its strength varies enormously in accordance with the positions of the itinerant electrons in the reciprocal space. As presented in Fig.~\ref{fig3}(d), the Heisenberg term mediated by itinerant electrons near M is vanishing small while it is finite corresponding to itinerant electrons near $\Gamma$. The phenomenon is closely related to the different forms of intersublattice coupling terms near the BZ center and corner.
For itinerant electrons near $\Gamma$ the two sublattices couple through the constant term $t^\prime s_0 \sigma_1$ as shown in Eq.~\eqref{2a eff G}, while at M the coupling term is $t^\prime k_x k_y s_0 \sigma_1$ as shown in Eq.~\eqref{2a eff H} which is vanishing small for $\kk$ near M.

The Ising and anisotropic terms are much weaker than the Heisenberg and DM terms. In Fig.~\ref{fig4}(a), we show the $T_{ij}$ corresponding to the condition where the two impurities are arranged along the $[11]$ direction in the same sublattice, i.e. $\RR \parallel [11]$. Along the $[11]$ direction, it satisfies $g_1 = g_2$ in Eq.~\eqref{range function2}, leading to that the Ising terms equals the anisotropic terms, i.e. $T_{xx} = T_{xy} = T_{yy}$.
In addition, these terms oscillate with two different periods $P_1$ and $P_2$ as indicated in Fig.~\ref{fig4} (a), which origin from the interference of two distinct Fermi surface as shown in Fig.~\ref{fig1}(b). These two periods can be estimated with $2\pi|k_{F1}+ k_{F2}|^{-1}=6.6$ and $2\pi|k_{F1}- k_{F2}|^{-1}=26$, with $k_{Fi}$ the $i$-th Fermi wave vectors along the $[11]$ direction \cite{zare2018strongly}.
In Fig.~\ref{fig4}(b), we show the anisotropic term $T_{xy}$ with respect to $\RR$ in different directions. As shown, $T_{xy}$ is prohibited along the $x$ ($y$) axis by the mirror symmetry $\{{M}_y|{\bf0}\}$ ($\{{M}_x|{\bf0}\}$).


\textit{Discussion and conclusion.}
In the above, based on the symmetry and numerical analysis, we show the rich and controllable RKKY interactions among impurities in the nonsymmorphic crystal respecting the $P4/nmm$ space group. The substantial DM interaction mediated by itinerant electrons near the BZ corner, whose form varies in accordance with the positions of the
magnetic impurities, can lead to rich magnetic spin textures in the system.
For instance, for impurities in the same sublattice at the $2a$ Wyckoff positions, the $D_{2d}$ symmetric DM interaction in Figs.~\ref{fig2}(a)(b) supports the anti-skyrmion type spin texture presented in Fig.~\ref{fig2}(c);
while for impurities in the same sublattice at the $2c$ Wyckoff positions, the $C_{4v}$ symmetric DM interaction in Figs.~\ref{fig2}(d)(e) favors the N$\acute{{\rm e}}$el-type skyrmion spin texture shown in Fig.~\ref{fig2}(f) \cite{koshibae2016theory,lin2021skyrmion,hayami2022skyrmion}.
In both cases, the skyrmion in the different sublattices carry opposite helicities \cite{doi:10.7566/JPSJ.89.013703,doi:10.7566/JPSJ.83.114704,doi:10.7566/JPSJ.85.124702,PhysRevB.102.195147}.
Moreover, the relative position of skyrmion centers in the two sublattices can be adjusted by the intersublattice Heisenberg interaction, which is weak mediated by itinerant electrons near the BZ corner as indicated in Fig.~\ref{fig3}(d). If the intersublattice interaction is strong, which can be true if there are additional itinerant electrons near the BZ center, the skyrmion spin textures in the two sublattices hybridize, and a spiral magnetic order may be supported instead \cite{lin2021skyrmion,hayami2022skyrmion}.

In summary, we theoretically investigate the RKKY interaction between magnetic impurities in a nonsymmorphic crystal respecting the space group $P4/nmm$. We show that though the system is globally centrosymmetric, various types of magnetic interactions, including the Heisenberg-type, the DM-type, the Ising-type and the anisotropic RKKY interactions, can appear according to the configurations of the impurities in the real space, and their strength can be controlled by adjusting the locations of the itinerant electrons in the reciprocal space. Our study reveals the role that the nonsymmorphic symmetries play in affecting magnetism, and suggests that the nonsymmorphic crystals are potential platforms to support rich types of magnetic orders.


\begin{acknowledgments}
The authors are grateful to Runze Chi and Chuhao Li for fruitful discussions in the numerical calculation of Green's function. This work is supported by the Ministry of Science and Technology of China 973 program (Grant No. 2017YFA0303100), National Science Foundation of China (Grant No. NSFC-12174428, NSFC-11888101 and NSFC-11920101005), and the Strategic Priority Research Program of Chinese Academy of Sciences (Grant No. XDB28000000  and No. XDB33000000).

\end{acknowledgments}


\begin{thebibliography}{10}
\bibitem{footnote_symm}
\bibinfo{title}{Here, we specify the point group part of those symmetries as
  $m_y:(x,y,z)\mapsto (x,-y,z)$, $s_{4z}:(x,y,z)\mapsto (y,-x,-z)$,
  $m_{xy}:(x,y,z)\mapsto (-y,-x,z)$, $c_{4z}:(x,y,z)\mapsto (-y,x,z)$,
  $i:(x,y,z)\mapsto (-x,-y,-z)$. The origin of the coordinate system is defined
  on the site in fig. 1}.
\bibitem{footnote1}
\bibinfo{title}{At the bz center, the fractional translation in a nonsymmorphic
  symmetry has no effect making the nonsymmorphic symmetry simply equivalent to
  the point group symmetry} .
\end{thebibliography}


\begin{widetext}

\renewcommand{\theequation}{A\arabic{equation}}
\renewcommand{\thefigure}{A\arabic{figure}}
\renewcommand{\thetable}{A\arabic{table}}
\setcounter{equation}{0}
\setcounter{figure}{0}
\setcounter{table}{0}

\section{APPENDIX A: The construction of effective model from $2c$ Wyckoff position}
In this section, we present the detailed construction of the low-energy effective model $\mathcal{H}_{\mathrm{M,2c}}$ near the boundary of the BZ, i.e., M point.
The bases in the reciprocal space and in the real space are related by the Fourier transform
\begin{eqnarray}\label{SM_Fourier}
| \phi_{\mathrm{2c,1}} ({\bf k}) \rangle = \sum_j e^{i {\bf k} \cdot {\bf R}_{\mathrm{2c,1}}^j } | \phi_{\mathrm{2c,1}} ({\bf R}_{\mathrm{2c,1}}^j) \rangle, \quad | \phi_{{\mathrm{2c,2}}} ({\bf k}) \rangle = \sum_j e^{i {\bf k} \cdot {\bf R}_{{\mathrm{2c,2}}}^j } | \phi_{{\mathrm{2c,2}}} ({\bf R}_{{\mathrm{2c,2}}}^j) \rangle,
\end{eqnarray}
where $\RR_{\mathrm{2c,i}}^j$ is the position of orbital located at the $i$-th 2c Wyckoff in the $j$-th unit cell satisfying $\RR_{\mathrm{2c,1}}^j-\RR_{{\mathrm{2c,2}}}^j=\ttau_0$, and $|\phi_{{\mathrm{2c,1/2}}}(\rr-\RR_{\mathrm{2c,1}}^j)\rangle$ are the localized atomic-like orbitals located in the position of $\RR_{\mathrm{2c,1}}^j/\RR_{\mathrm{2c,2}}^j$.

Firstly, we need to figure out how the symmetry operations act on the basis $(|\phi_{\mathrm{2c,1}}^\kk(\rr)\rangle,|\phi_{\mathrm{2c,2}}^\kk(\rr)\rangle)=(c_{{\mathrm{2c,1}}}^{\kk\dagger},c_{{\mathrm{2c,2}}}^{\kk\dagger})|0\rangle$, where $|0\rangle$ is the vacuum state, $c_{{\mathrm{2c}},i}^{\kk\dagger}$ is the creation operator of the orbitals located at $i$-th 2c Wyckoff position, and the spin index has been omitted for convenience.
When a spatial symmetry $\{\hat{g}|\bma_i\}$ acts on the basis function, we have
\begin{equation}\label{symmetry_matrix_convention_I}
\begin{aligned}
\{\hat{g}|0\}|\phi_{\mathrm{2c,1}}^\kk(\rr)\rangle&=|\phi_{\mathrm{2c,1}}^\kk(g^{-1}\rr)\rangle=\eta_{\mathrm{2c,1}}|\phi_{\mathrm{2c,1}}^\kk(\rr)\rangle,\\
\{\hat{g}|0\}|\phi_{\mathrm{2c,2}}^\kk(\rr)\rangle&=|\phi_{\mathrm{2c,2}}^\kk(g^{-1}\rr)\rangle=\sum_{j}e^{i\kk\cdot(\RR^j+\ttau_0)}|\phi_{\mathrm{2c,2}}(g^{-1}[\rr-\RR ^j-\ttau_0])\rangle\\
&=\sum_{j}e^{i\kk\cdot(\RR^j+\ttau_0)}\eta_{\mathrm{2c,2}}|\phi_{\mathrm{2c,2}}(\rr-g^{-1}\RR ^j-g^{-1}\ttau_0)\rangle\\
&=\eta_{\mathrm{2c,2}}\sum_{j}e^{i\kk\cdot[(\RR^j+\ttau_0)-(g^{-1}\RR ^j+g^{-1}\ttau_0)]}e^{i\kk\cdot(g^{-1}\RR ^j+g^{-1}\ttau_0)}|\phi_{\mathrm{2c,2}}(\rr-g^{-1}\RR ^j-g^{-1}\ttau_0)\rangle\\
&=\eta_{\mathrm{2c,2}}\sum_{j}e^{-i(g\kk-\kk)\cdot\RR^j}e^{-i(g\kk-\kk)\cdot\ttau_0}e^{i\kk\cdot(g^{-1}\RR ^j+g^{-1}\ttau_0)}|\phi_{\mathrm{2c,2}}(\rr-g^{-1}\RR ^j-g^{-1}\ttau_0)\rangle\\
&=\eta_{\mathrm{2c,2}}\sum_{j} e^{-i(g\kk-\kk)\cdot\ttau_0}e^{i\kk\cdot(g^{-1}\RR ^j+g^{-1}\ttau_0)}|\phi_{\mathrm{2c,2}}(\rr-g^{-1}\RR ^j-g^{-1}\ttau_0)\rangle\\
&=\eta_{\mathrm{2c,2}}e^{-i(g\kk-\kk)\cdot\ttau_0}\sum_{j} e^{i\kk\cdot(g^{-1}\RR ^j+g^{-1}\ttau_0)}|\phi_{\mathrm{2c,2}}(\rr-g^{-1}\RR ^j-g^{-1}\ttau_0)\rangle=\eta_{\mathrm{2c,2}}e^{i(\kk-g\kk)\cdot\ttau_0}|\phi_{\mathrm{2c,2}}^\kk(\rr)\rangle,\\
\{E|\bma_i\}|\phi_{\mathrm{2c,1}}^\kk(\rr)\rangle&=e^{-i\kk\cdot\bma_i}|\phi_{\mathrm{2c,1}}^\kk(\rr)\rangle,\\
\{E|\bma_i\}|\phi_{\mathrm{2c,2}}^\kk(\rr)\rangle&=e^{-i\kk\cdot\bma_i}|\phi_{\mathrm{2c,2}}^\kk(\rr)\rangle,\\
\{E|\ttau_0\}|\phi_{\mathrm{2c,1}}^\kk(\rr)\rangle& =|\phi_{\mathrm{2c,2}}^\kk(T_{\ttau_0}^{-1}\rr)\rangle=\sum_{j}e^{i\kk\cdot(\RR^j+\ttau_0-\ttau_0)}|\phi_{\mathrm{2c,2}}( \rr-\RR ^j-\ttau_0)\rangle=e^{-i\kk\cdot\ttau_0}|\phi_{\mathrm{2c,2}}^\kk(\rr)\rangle\\
\{E|\ttau_0\}|\phi_{\mathrm{2c,2}}^\kk(\rr)\rangle& =|\phi_{\mathrm{2c,1}}^\kk(T_{\ttau_0}^{-1}\rr)\rangle=\sum_{j}e^{i\kk\cdot(\RR^j+\ttau_0+\ttau_0-\ttau_0)}|\phi_{\mathrm{2c,1}}( \rr-\RR ^j-\ttau_0-\ttau_0)\rangle=e^{-i\kk\cdot\ttau_0}|\phi_{\mathrm{2c,1}}^\kk(\rr)\rangle
\end{aligned}
\end{equation}
where $\eta_{\mathrm{2c,i}}$ is the $\hat{g}$'s eigenvalue of the orbital located at $i$-th 2c Wyckoff position.
Thus, the matrices of the generator of the little group $G_{\kk}$ at the $\Gamma$ and M points are summarized in Table~\ref{symmetry_matrix_C4v_M}.
\begin{table}[h]
\caption{\label{symmetry_matrix_C4v_M} { Matrix form for the symmetry operations, $\{ I | {\bf \tau_0} \}$, $\{ M_y | {\bf 0} \}$, $\{ C_{4z} | {\bf 0} \}$ and $\mathcal{T}$, at $\Gamma$ and M point. } }
\begin{tabular}{ p{1.8cm}<{\centering} p{1.8cm}<{\centering} p{1.8cm}<{\centering} p{1.8cm}<{\centering} p{1.8cm}<{\centering} }  \hline\hline
           &   $\{ I | {\bf \tau_0} \}$   &  $\{ M_y | {\bf 0} \}$   &   $\{ C_{4z} | {\bf 0} \}$                 &   $\mathcal{T}$      \\ \hline
$\Gamma$  &  $  -s_0 \sigma_1$  &  $-i s_2 \sigma_0$  &  $e^{-is_3\pi/4 } \sigma_0$   &  $i s_2 \sigma_0 K$    \\
M  &  $ -s_0 \sigma_1$  &  $-i s_2 \sigma_3$  &  $e^{-is_3\pi/4 } \sigma_3$   &  $i s_2 \sigma_0 K$     \\ \hline
\end{tabular}
\end{table}

Due to the fact that the single-particle Hamitonian $H(\kk)=f_{ij}(\kk)s_i\sigma_j$ is a bilinear map on the single-particle Hilbert space, the matrix form and $f(\kk)$ can furnish a representation of $D_{4h}$.
We first consider the time reversal symmetry and the inversion symmetry, which constrain the system as
\begin{eqnarray}\label{SM_constrain_kp}
\mathcal{T} \mathcal{H}_{\mathrm{\Gamma/M,2c}}({\bf k}) \mathcal{T}^{-1} = \mathcal{H}_{\mathrm{\Gamma/M,2c}}(-{\bf k}), \quad
\mathcal{I} \mathcal{H}_{\mathrm{\Gamma/M,2c}}({\bf k}) \mathcal{I}^{-1} = \mathcal{H}_{\mathrm{\Gamma/M,2c}}(-{\bf k}).
\end{eqnarray}
The four-band model $\mathcal{H}_{{\mathrm{\Gamma/M,2c}}}({\bf k})$ can be generally expressed in the form of the sixteen $\Gamma = s_i \sigma_j$ matrices. The constraints in Eq.~\eqref{SM_constrain_kp} merely allow six $\Gamma$ matrices, i.e. $s_0 \sigma_0$, $s_0 \sigma_1$, $s_0 \sigma_2$, $s_1 \sigma_3$, $s_2 \sigma_3$ and $s_3 \sigma_3$, to appear in $\mathcal{H}_{{\mathrm{\Gamma/M,2c}}}({\bf k})$.
Then, we consider the constraints of the crystalline symmetries.
At the $\Gamma$ point of BZ, one can classify the above six matrices as shown in Table.\ref{symmetry_rep_C4v_G}.
\begin{table}[htbp!]
\caption{\label{symmetry_rep_C4v_G} { Classification table of the matrix and $f((\pi,\pi)+\kk)$ according the $D_{4h}$. } }
\begin{tabular}{ p{1.8cm}<{\centering} p{1.8cm}<{\centering} p{1.8cm}<{\centering} p{1.8cm}<{\centering} p{1.8cm}<{\centering} p{1.8cm}<{\centering} p{1.8cm}<{\centering}p{1.8cm}<{\centering} }  \hline\hline
         IRs  &   $\{ E | {\bf 0} \}$ &   $\{ I | {\bf \tau_0} \}$  &   $\{ C_{4z} | {\bf 0} \}$ &  $\{ M_y | {\bf 0} \}$                    &   $s_i\sigma_j$ & $f_{ij}(\kk)$      \\ \hline
         $A_{1g}$  & $1$ & $1$  & $1$  & $1$  & $s_0\sigma_0,s_0\sigma_1$& $\text{cons.,}k_x^2+k_y^2$\\ \hline
         $A_{1u} $       & $1$ & $-1$ & $1$  & $-1$ & $s_3\sigma_3$ &$o(k^2)$\\ \hline
         $A_{2u} $       & $1$ & $-1$ & $1$ & $1$ & $s_0\sigma_2$ &  $o(k^2)$ \\ \hline
         $E_{u} $       & $2$ & $-2$ & $0$  & $0$  & $(s_2\sigma_3,-s_1\sigma_3)$ &$(k_x,k_y)$\\ \hline
\end{tabular}
\end{table}
Therefore, $\mathcal{H}_{\mathrm{\Gamma,2c}}({\bf k})$ must take the following form
\begin{eqnarray}\label{SM_G_kp}
\mathcal{H}_{\mathrm{\Gamma,2c}}({\bf k}) = (t(k_x^2+k_y^2)-\mu)s_0\sigma_0+\lambda(k_xs_2\sigma_3-k_ys_1\sigma_3)+t^\prime s_0\sigma_1.
\end{eqnarray}
The corresponding energy dispersions and the spin-polarization in the $A$-sublattice space are shown in Fig.~\ref{SM energy dispersions}(a).
It can be clearly seen that the spin polarization near the $\Gamma$ point is almost 0.
Note that the length of arrow representing the strength of spin-polarization in Fig.~\ref{SM energy dispersions}(a) is magnified by a factor of 20.
\begin{figure}[htbp!]
  \centering
  \includegraphics[width=0.47\linewidth]{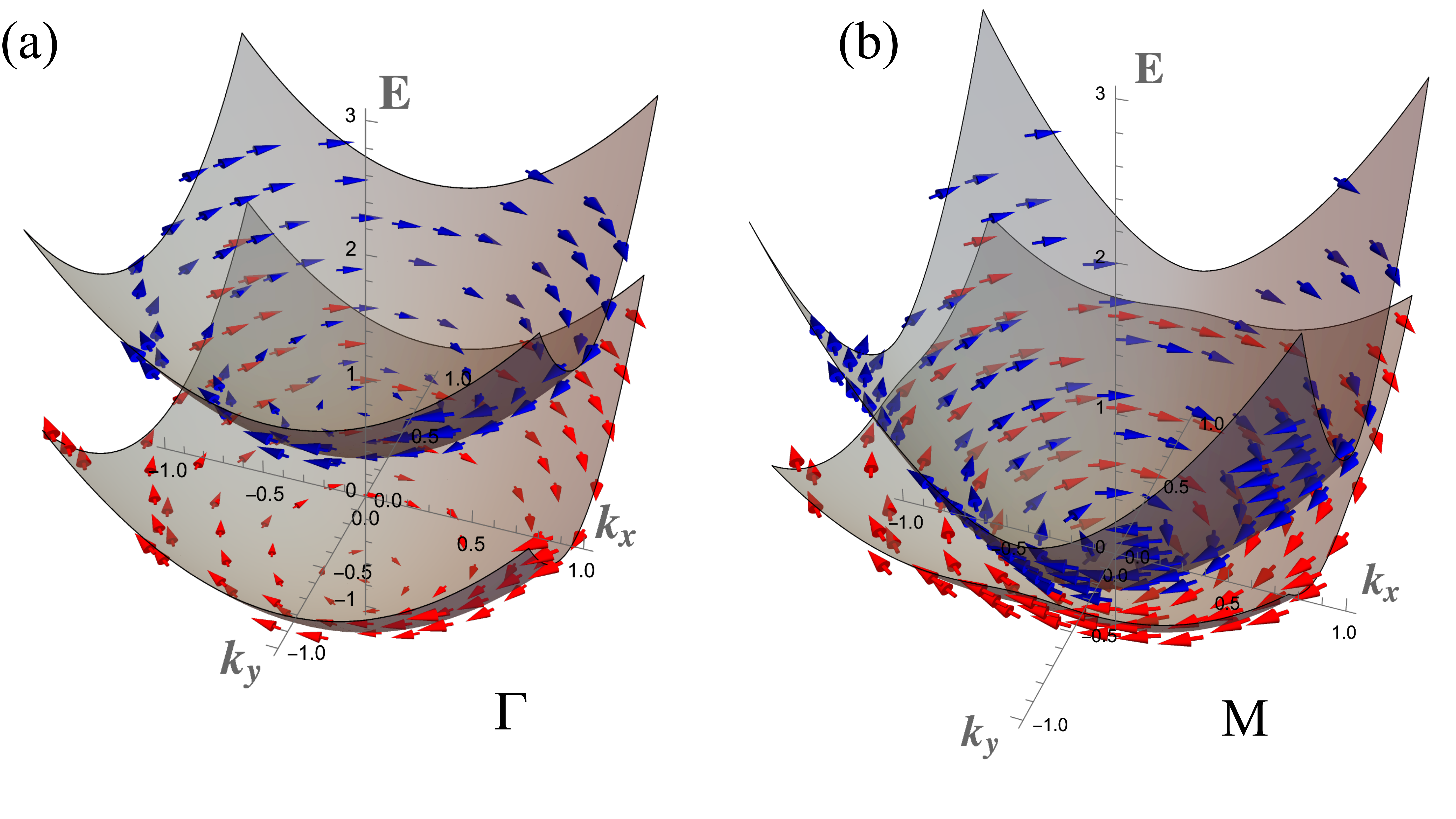}
  \caption{\label{SM energy dispersions}(a) The band structure near the M point plotted from the Hamiltonian in Eq.~(\ref{SM_M_kp}) with parameters $\{t, t_0, \lambda\} = \{1.0, 0.7, 0.15\}$.
  (b) the the band structure near the $\Gamma$ point, plotted from the Hamiltonian in Eq.~(\ref{SM_G_kp}) with the same parameters as (a).
  The length of the arrow represents the strength $\langle\bms_{A}\rangle$ of the spin polarization.
  In (a), the length of the arrow is magnified by a factor of 20.}
\end{figure}

Similarly, at the M point of BZ, one can classify the above six matrices as shown in Table.\ref{symmetry_rep_C4v_M}.
\begin{table}[htbp!]
\caption{\label{symmetry_rep_C4v_M} { Classification table of the matrix and $f( \kk)$ according the $D_{4h}$. } }
\begin{tabular}{ p{1.8cm}<{\centering} p{1.8cm}<{\centering} p{1.8cm}<{\centering} p{1.8cm}<{\centering} p{1.8cm}<{\centering} p{1.8cm}<{\centering} p{1.8cm}<{\centering}p{1.8cm}<{\centering} }  \hline\hline
         IRs  &   $\{ E | {\bf 0} \}$ &   $\{ I | {\bf \tau_0} \}$  &   $\{ C_{4z} | {\bf 0} \}$ &  $\{ M_y | {\bf 0} \}$                    &   $s_i\sigma_j$ & $f_{ij}(\kk)$      \\ \hline
         $A_{1g}$ & $1$ & $1$  & $1$  & $1$  & $s_0\sigma_0$& $\text{cons.,}k_x^2+k_y^2$ \\ \hline
         $B_{2g}$ & $1$ & $1$ & $-1$ & $-1$ & $s_0\sigma_1$ & $k_xk_y $  \\ \hline
         $A_{1u}$ & $1$ & $-1$ & $1$  & $-1$ & $s_3\sigma_3$ & $o(k^2) $  \\ \hline
         $B_{1u}$ & $1$ & $-1$ & $-1$ & $-1$ & $s_0\sigma_2$ & $o(k^2) $  \\ \hline
         $E_u$    & $2$ & $-2$ & $0$  & $0$  & $(s_2\sigma_3,-s_1\sigma_3)$ & $ (k_x,k_y)$ \\ \hline
\end{tabular}
\end{table}
Therefore, $\mathcal{H}_{\mathrm{M,2c}}({\bf k})$ must take the following form
\begin{eqnarray}\label{SM_M_kp}
\mathcal{H}_{\mathrm{M,2c}}({\bf k}) = (t(k_x^2+k_y^2)-\mu)s_0\sigma_0+\lambda(k_xs_2\sigma_3-k_ys_1\sigma_3)+t^\prime .
\end{eqnarray}
The corresponding energy dispersions and the spin-polarization in the $A$-sublattice space are shown in Fig.~\ref{SM energy dispersions}(b).

\renewcommand{\theequation}{B\arabic{equation}}
\renewcommand{\thefigure}{B\arabic{figure}}
\renewcommand{\thetable}{B\arabic{table}}
\setcounter{equation}{0}
\setcounter{figure}{0}
\setcounter{table}{0}

\section{APPENDIX B: The Green's function in the momentum-energy space}
In this section, we give a detailed derivation of the Green's function in the momentum-energy space.
The low-energy effective model $H_{\mathrm{2a,M}}$ is
\begin{equation}\label{effective model 2a at M point}
\mathcal{H}_{\mathrm{M,2a}}(\kk)=m(\kk)s_0\sigma_0+\lambda k_x s_2\sigma_3+\lambda k_y s_1\sigma_3+t^\prime k_xk_y s_0\sigma_1.
\end{equation}
where $m(\kk)=t(k_x^2+k_y^2)$ with $t,t^\prime,\lambda$ all constants.
The bases in the reciprocal space and in the real space are related by the Fourier transform
\begin{eqnarray}\label{SM_Fourier}
| \phi_{\mathrm{2a,1}} ({\bf k}) \rangle = \sum_j e^{i {\bf k} \cdot {\bf R}_{\mathrm{2a,1}}^j } | \phi_{\mathrm{2a,1}} ({\bf R}_{\mathrm{2a,1}}^j) \rangle, \quad | \phi_{{\mathrm{2a,2}}} ({\bf k}) \rangle = \sum_j e^{i {\bf k} \cdot {\bf R}_{{\mathrm{2a,2}}}^j } | \phi_{{\mathrm{2a,2}}} ({\bf R}_{{\mathrm{2a,2}}}^j) \rangle,
\end{eqnarray}
where $\RR_{\mathrm{2a,i}}^j$ is the position of orbital located at the $i$-th 2c Wyckoff in the $j$-th unit cell satisfying $\RR_{\mathrm{2a,1}}^j-\RR_{{\mathrm{2a,2}}}^j=\ttau_0$, and $|\phi_{{\mathrm{2a,1/2}}}(\rr-\RR_{\mathrm{2a,1}}^j)\rangle$ are the localized atomic-like orbitals located in the position of $\RR_{\mathrm{2a,1}}^j/\RR_{\mathrm{2a,2}}^j$.
The energy dispersions are
\begin{equation}
E_\pm(k_x,k_y)=t(k_x^2+k_y^2)\pm\sqrt{\lambda^2(k_x^2+k_y^2)+t^{\prime2}k_x^2k_y^2}\equiv m(\kk)\pm\Xi(\kk),
\end{equation}
which are two-fold degenerate.
The corresponding degenerate eigenstates $|\Psi_{\pm}(\kk)\rangle$ are
\begin{equation}\label{conduction electrons wavefunction}
\begin{aligned}
&|\Psi_{+(1)}(\kk)\rangle=\frac{1}{\sqrt{2}}[\frac{\lambda(-ik_x+k_y)}{\Xi(\kk)},-1,0 , \frac{t^\prime k_xk_y}{\Xi(\kk)}]^T,\
&|\Psi^\prime_{+(2)}(\kk)\rangle= \frac{1}{\sqrt{2}}[\frac{-t^\prime k_xk_y}{\Xi(\kk)} ,0 ,1 , \frac{\lambda(ik_x+k_y)}{\Xi(\kk)}]^T\\
&|\Psi_{-(3)}(\kk)\rangle= \frac{1}{\sqrt{2}}[\frac{\lambda(-ik_x+k_y)}{\Xi(\kk)},1,0 , \frac{t^\prime k_xk_y}{\Xi(\kk)}]^T,\
&|\Psi^\prime_{-(4)}(\kk)\rangle=\frac{1}{\sqrt{2}}[\frac{t^\prime k_xk_y}{\Xi(\kk)} ,0 ,1 , \frac{-\lambda(ik_x+k_y)}{\Xi(\kk)}]^T.
\end{aligned}
\end{equation}

In order to obtain the retarded Green's function in real space, we first calculate the corresponding Green's function in momentum space:
\begin{equation}
[G^R_0(\kk,\epsilon)]_{\mu\mu^\prime}=\sum_{n=1}^{4}\frac{|\Psi_{n}(\kk)\rangle_\mu{}_{\mu^\prime}\langle\Psi_{n}(\kk)|}{\epsilon-E_n(\kk)+i\eta^+}\Rightarrow [G_0^R(\RR_i,\RR_j,\epsilon)]_{\mu\mu^\prime}=\int \frac{\mathrm{d}^2\kk}{(2\pi)^2}\langle \RR_i|[G^R_0(\kk,\epsilon)]_{\mu\mu^\prime}|\RR_j\rangle ,
\end{equation}
where $\kk=(k_x,k_y)$. After some straightforward calculation, we have
\begin{align}
G_{\mathrm{M,2a}}^R(\kk,\epsilon)=\frac{1}{2}\frac{1}{\epsilon-E_-(\kk)+i\eta^+}[s_0 \sigma_0- \frac{t^\prime k_x k_y}{\Xi(\kk)} s_0 \sigma_1 - \frac{\lambda k_y}{\Xi(\kk)} s_1 \sigma_3-\frac{\lambda k_x}{\Xi(\kk)} s_2 \sigma_3   ]\\
+\frac{1}{2}\frac{1}{\epsilon-E_+(\kk)+i\eta^+}[s_0 \sigma_0+ \frac{t^\prime k_x k_y}{\Xi(\kk)} s_0 \sigma_1 + \frac{\lambda k_y}{\Xi(\kk)} s_1 \sigma_3+\frac{\lambda k_x}{\Xi(\kk)} s_2 \sigma_3   ],
\end{align}
where $\Xi(\kk)=\sqrt{\lambda^2(k_x^2+k_y^2)+t^{\prime2}k_x^2k_y^2}$, $E_\pm(k_x,k_y)= m(\kk)\pm\Xi(\kk)$.
The real-space Green's function can be obtained as
\begin{equation}
\begin{aligned}
&G_{\mathrm{M,2a}}(\RR_1,\RR_2,\epsilon)=\ g_0 s_0\sigma_0+g_1 s_2\sigma_3+g_2 s_1\sigma_3+g^\prime s_0\sigma_1,\\
\end{aligned}
\end{equation}
where $g_0,g_1,g_2,g^\prime$ are the $(\RR_1-\RR_2)$- and $\omega$- dependent functions:
\begin{equation}
\begin{aligned}
&g_0 =\frac{1}{2}\sum_{\alpha=\pm}\int \frac{\mathrm{d}^2\kk}{(2\pi)^2} e^{i\kk\cdot(\RR_1-\RR_2)}\frac{1}{\Xi(\kk)}\frac{1}{\omega-E_{\alpha}(\kk)+i0^+},\qquad g_1 =\frac{1}{2}\sum_{\alpha=\pm}\int \frac{\mathrm{d}^2\kk}{(2\pi)^2} e^{i\kk\cdot(\RR_1-\RR_2)}\frac{1}{\Xi(\kk)}\frac{\alpha\lambda k_x}{\omega-E_{\alpha}(\kk)+i0^+},\\
&g_2 =\frac{1}{2}\sum_{\alpha=\pm}\int \frac{\mathrm{d}^2\kk}{(2\pi)^2} e^{i\kk\cdot(\RR_1-\RR_2)}\frac{1}{\Xi(\kk)}\frac{\alpha\lambda k_y}{\omega-E_{\alpha}(\kk)+i0^+},\qquad g^\prime =\frac{1}{2}\sum_{\alpha=\pm}\int \frac{\mathrm{d}^2\kk}{(2\pi)^2} e^{i\kk\cdot(\RR_1-\RR_2+\ttau_0)}\frac{1}{\Xi(\kk)}\frac{\alpha t^\prime k_xk_y}{\omega-E_{\alpha}(\kk)+i0^+}.
\end{aligned}
\end{equation}

Similarly, for the effective model $\mathcal{H}_{\mathrm{M,2c}}$, the corresponding Green's function is
\begin{equation}
\begin{aligned}
G_{\mathrm{M,2c}}^R(\kk,\epsilon)=\frac{1}{2}\frac{1}{\epsilon-E_-(\kk)+i\eta^+}[s_0 \sigma_0- \frac{t^\prime k_x k_y}{\Xi(\kk)} s_0 \sigma_1 + \frac{\lambda k_y}{\Xi(\kk)} s_1 \sigma_3-\frac{\lambda k_x}{\Xi(\kk)} s_2 \sigma_3   ]\\
+\frac{1}{2}\frac{1}{\epsilon-E_+(\kk)+i\eta^+}[s_0 \sigma_0+ \frac{t^\prime k_x k_y}{\Xi(\kk)} s_0 \sigma_1 - \frac{\lambda k_y}{\Xi(\kk)} s_1 \sigma_3+\frac{\lambda k_x}{\Xi(\kk)} s_2 \sigma_3   ],
\end{aligned}
\end{equation}
where $\Xi(\kk)=\sqrt{\lambda^2(k_x^2+k_y^2)+t^{\prime2}k_x^2k_y^2}$, $E_\pm(k_x,k_y)= m(\kk)\pm\Xi(\kk)$.
The real-space Green's function can be obtained as
\begin{equation}
\begin{aligned}
&G_{\mathrm{M,2c}}(\RR_1,\RR_2,\epsilon)=\ g_0 s_0\sigma_0+g_1 s_2\sigma_3+g_2 s_1\sigma_3+g^\prime s_0\sigma_1,\\
\end{aligned}
\end{equation}
where $g_0,g_1,g_2,g^\prime$ are the $(\RR_1-\RR_2)$- and $\omega$- dependent functions:
\begin{equation}
\begin{aligned}
&g_0 =\frac{1}{2}\sum_{\alpha=\pm}\int \frac{\mathrm{d}^2\kk}{(2\pi)^2} e^{i\kk\cdot(\RR_1-\RR_2)}\frac{1}{\Xi(\kk)}\frac{1}{\omega-E_{\alpha}(\kk)+i0^+},\qquad g_1 =\frac{1}{2}\sum_{\alpha=\pm}\int \frac{\mathrm{d}^2\kk}{(2\pi)^2} e^{i\kk\cdot(\RR_1-\RR_2)}\frac{1}{\Xi(\kk)}\frac{\alpha\lambda k_x}{\omega-E_{\alpha}(\kk)+i0^+},\\
&g_2 =\frac{1}{2}\sum_{\alpha=\pm}\int \frac{\mathrm{d}^2\kk}{(2\pi)^2} e^{i\kk\cdot(\RR_1-\RR_2)}\frac{1}{\Xi(\kk)}\frac{-\alpha\lambda k_y}{\omega-E_{\alpha}(\kk)+i0^+},\qquad g^\prime =\frac{1}{2}\sum_{\alpha=\pm}\int \frac{\mathrm{d}^2\kk}{(2\pi)^2} e^{i\kk\cdot(\RR_1-\RR_2+\ttau_0)}\frac{1}{\Xi(\kk)}\frac{\alpha t^\prime k_xk_y}{\omega-E_{\alpha}(\kk)+i0^+}.\\
\end{aligned}
\end{equation}

For the effective model $\mathcal{H}_{\mathrm{\Gamma,2a}}$, the corresponding Green's function is
\begin{align}
G_{\mathrm{\Gamma,2a}}^R(\kk,\epsilon)=\frac{1}{2}\frac{1}{\epsilon-E^\prime_-(\kk)+i\eta^+}[s_0 \sigma_0- \frac{t^\prime}{\Xi^\prime (\kk)} s_0 \sigma_1 - \frac{\lambda k_y}{\Xi^\prime (\kk)} s_1 \sigma_3-\frac{\lambda k_x}{\Xi^\prime (\kk)} s_2 \sigma_3   ]\\
+\frac{1}{2}\frac{1}{\epsilon-E^\prime_+(\kk)+i\eta^+}[s_0 \sigma_0+ \frac{t^\prime}{\Xi^\prime (\kk)} s_0 \sigma_1 + \frac{\lambda k_y}{\Xi^\prime (\kk)} s_1 \sigma_3+\frac{\lambda k_x}{\Xi^\prime (\kk)} s_2 \sigma_3   ],
\end{align}
where $\Xi(\kk)=\sqrt{\lambda^2(k_x^2+k_y^2)+t^{\prime2}}$, $E_\pm(k_x,k_y)= m(\kk)\pm\Xi(\kk)$.
The real-space Green's function can be obtained as
\begin{equation}
\begin{aligned}
&G_{\mathrm{\Gamma,2a}}(\RR_1,\RR_2,\epsilon)=\ g_0 s_0\sigma_0+g_1 s_2\sigma_3+g_2 s_1\sigma_3+g^\prime s_0\sigma_1,\\
\end{aligned}
\end{equation}
where $g_0,g_1,g_2,g^\prime$ are the $(\RR_1-\RR_2)$- and $\omega$- dependent functions:
\begin{equation}
\begin{aligned}
&g_0 =\frac{1}{2}\sum_{\alpha=\pm}\int \frac{\mathrm{d}^2\kk}{(2\pi)^2} e^{i\kk\cdot(\RR_1-\RR_2)}\frac{1}{\Xi(\kk)}\frac{1}{\omega-E_{\alpha}(\kk)+i0^+},\qquad g_1 =\frac{1}{2}\sum_{\alpha=\pm}\int \frac{\mathrm{d}^2\kk}{(2\pi)^2} e^{i\kk\cdot(\RR_1-\RR_2)}\frac{1}{\Xi(\kk)}\frac{\alpha\lambda k_x}{\omega-E_{\alpha}(\kk)+i0^+},\\
&g_2 =\frac{1}{2}\sum_{\alpha=\pm}\int \frac{\mathrm{d}^2\kk}{(2\pi)^2} e^{i\kk\cdot(\RR_1-\RR_2)}\frac{1}{\Xi(\kk)}\frac{\alpha\lambda k_y}{\omega-E_{\alpha}(\kk)+i0^+},\qquad g^\prime =\frac{1}{2}\sum_{\alpha=\pm}\int \frac{\mathrm{d}^2\kk}{(2\pi)^2} e^{i\kk\cdot(\RR_1-\RR_2+\ttau_0)}\frac{1}{\Xi(\kk)}\frac{\alpha t^\prime  }{\omega-E_{\alpha}(\kk)+i0^+}.\\
\end{aligned}
\end{equation}

For the effective model $\mathcal{H}_{\mathrm{\Gamma,2a}}$, the corresponding Green's function is
\begin{align}
G_{\mathrm{\Gamma,2c}}^R(\kk,\epsilon)=\frac{1}{2}\frac{1}{\epsilon-E^\prime_-(\kk)+i\eta^+}[s_0 \sigma_0- \frac{t^\prime}{\Xi^\prime (\kk)} s_0 \sigma_1 + \frac{\lambda k_y}{\Xi^\prime (\kk)} s_1 \sigma_3-\frac{\lambda k_x}{\Xi^\prime (\kk)} s_2 \sigma_3   ]\\
+\frac{1}{2}\frac{1}{\epsilon-E^\prime_+(\kk)+i\eta^+}[s_0 \sigma_0+ \frac{t^\prime}{\Xi^\prime (\kk)} s_0 \sigma_1 - \frac{\lambda k_y}{\Xi^\prime (\kk)} s_1 \sigma_3+\frac{\lambda k_x}{\Xi^\prime (\kk)} s_2 \sigma_3   ],
\end{align}
where $\Xi(\kk)=\sqrt{\lambda^2(k_x^2+k_y^2)+t^{\prime2}}$, $E_\pm(k_x,k_y)= m(\kk)\pm\Xi(\kk)$.
The real-space Green's function can be obtained as
\begin{equation}
\begin{aligned}
&G_{\mathrm{\Gamma,2c}}(\RR_1,\RR_2,\epsilon)=\ g_0 s_0\sigma_0+g_1 s_2\sigma_3+g_2 s_1\sigma_3+g^\prime s_0\sigma_1,\\
\end{aligned}
\end{equation}
where $g_0,g_1,g_2,g^\prime$ are the $(\RR_1-\RR_2)$- and $\omega$- dependent functions:
\begin{equation}
\begin{aligned}
&g_0 =\frac{1}{2}\sum_{\alpha=\pm}\int \frac{\mathrm{d}^2\kk}{(2\pi)^2} e^{i\kk\cdot(\RR_1-\RR_2)}\frac{1}{\Xi(\kk)}\frac{1}{\omega-E_{\alpha}(\kk)+i0^+},\qquad g_1 =\frac{1}{2}\sum_{\alpha=\pm}\int \frac{\mathrm{d}^2\kk}{(2\pi)^2} e^{i\kk\cdot(\RR_1-\RR_2)}\frac{1}{\Xi(\kk)}\frac{\alpha\lambda k_x}{\omega-E_{\alpha}(\kk)+i0^+},\\
&g_2 =\frac{1}{2}\sum_{\alpha=\pm}\int \frac{\mathrm{d}^2\kk}{(2\pi)^2} e^{i\kk\cdot(\RR_1-\RR_2)}\frac{1}{\Xi(\kk)}\frac{-\alpha\lambda k_y}{\omega-E_{\alpha}(\kk)+i0^+},\qquad g^\prime =\frac{1}{2}\sum_{\alpha=\pm}\int \frac{\mathrm{d}^2\kk}{(2\pi)^2} e^{i\kk\cdot(\RR_1-\RR_2+\ttau_0)}\frac{1}{\Xi(\kk)}\frac{\alpha t^\prime  }{\omega-E_{\alpha}(\kk)+i0^+}.\\
\end{aligned}
\end{equation}

\renewcommand{\theequation}{C\arabic{equation}}
\renewcommand{\thefigure}{C\arabic{figure}}
\renewcommand{\thetable}{C\arabic{table}}
\setcounter{equation}{0}
\setcounter{figure}{0}
\setcounter{table}{0}

\section{APPENDIX C: The derivation of the RKKY interaction from the Green's function}
In this section, we give the detailed derivation of the RKKY interaction from the real-space Green's function.
In the above section, we show that the real-space Green's function has the form:
\begin{equation}
G_0(\RR_1,\RR_2,\epsilon)=\ g_0 s_0\sigma_0+g_1 s_2\sigma_3+g_2 s_1\sigma_3+g^\prime s_0\sigma_1,
\end{equation}
where $g_i$ depends on the loactions of the Fermi surface and the orbitals.

Firstly, we consider the RKKY interaction between the same sublattice $A$, which is given by
\begin{equation}\label{SM_RKKY_AA}
\begin{aligned}
H^A_{\mathrm{RKKY}}&=-\frac{J^2}{\pi}\mathrm{Im} \int_{\epsilon <\mu} \mathrm{d}\epsilon\ \mathrm{Tr} [\bmS_2\cdot[\frac{\sigma_0+\sigma_z}{2}\otimes \bms] G_{0}^R(\RR_2,\RR_1,\epsilon )\bmS_1\cdot[\frac{\sigma_0+\sigma_z}{2}\otimes \bms] G_{0}^R(\RR_1,\RR_2,\epsilon )]\\
&=-\frac{J^2}{\pi}\mathrm{Im}\int_{\epsilon <\mu} \mathrm{d}\epsilon\ \mathrm{Tr} [\bmS_2\cdot  \bms G_{0,A}^R(\RR_2,\RR_1,\epsilon )\bmS_1\cdot\bms G_{0,A}^R(\RR_1,\RR_2,\epsilon )]
\end{aligned}
\end{equation}
where $G_{0,A}^R(\RR_2,\RR_1,\epsilon)$ is defined as
\begin{equation}
G_{0,A}^R(\RR_1,\RR_2,\epsilon)\equiv\langle \sigma_z=+1|G_{0}(\RR_1,\RR_2,\epsilon)|\sigma_z=+1\rangle=\ g_0 s_0 +g_1 s_2 +g_2 s_1.
\end{equation}
The integrand of Eq.~(\ref{SM_RKKY_AA}) can be expanded as
\begin{equation}\label{SM_RKKY_AA_expand}
\sum_{\alpha,\beta=1}^3S_{2\alpha}S_{1\beta}{\mathrm{Tr}}[s_\alpha (g_0 s_0-g_2s_1-g_1s_2) s_\beta (g_0s_0+g_2s_1+g_1s_2)],
\end{equation}
where we use the fact that $g_0(\RR_1,\RR_2)=g_0(\RR_2,\RR_1)$, $g_1(\RR_1,\RR_2)=-g_1(\RR_2,\RR_1)$, and $g_2(\RR_1,\RR_2)=-g_2(\RR_2,\RR_1)$.
The Eq.~(\ref{SM_RKKY_AA_expand}) contains five terms:
\begin{equation}
(g_1^2+g_2^2+g_0^2) \bmS_{1}\cdot\bmS_{2},\ 2ig_2g_0(\bmS_{1 }\times \bmS_{2 })_{x},\ 2ig_1g_0(\bmS_{1 }\times \bmS_{2 })_{y},-2g_2^2S_{1x}S_{2x},\ -2g_1^2S_{1y}S_{2y},\ -2g_1g_2(S_{1 x}S_{2 y}+S_{1y}S_{2 x}).
\end{equation}
Finally, the RKKY interaction $H^A_{\mathrm{RKKY}}$ between the same sublattice A can be divided into four type: the Heisenberg-type $\Lambda(\RR,\mu)\bmS_{1}\cdot\bmS_{2}$, the DM-type $\mathrm{D}_{i}(\bmS_{1 }\times \bmS_{2 })_i$, the Ising-type ($T_{ii}S_{1i}S_{2i}$) and the anisotropic terms ($T_{i \neq j}S_{1i}S_{2j}$), of which strength can be calculated by
\begin{equation}\label{SM range function1}
\begin{aligned}
\Lambda(\RR,\mu)&=-\frac{J^2}{\pi}\mathrm{Im}\int_{\omega <\mu} \mathrm{d}\omega\ (g_0^2+g_1^2+g_2^2),\qquad \mathrm{D}_{1}(\RR,\mu) =-\frac{2 J^2}{\pi}\mathrm{Re}\int_{\omega <\mu} \mathrm{d}\omega\ g_0g_2,\\
\mathrm{D}_{2}(\RR,\mu)&=-\frac{2 J^2}{\pi}\mathrm{Re}\int_{\omega <\mu} \mathrm{d}\omega\ g_0g_1,\qquad
T_{ij}(\RR,\mu) =-\frac{J^2}{\pi}\mathrm{Im}\int_{\omega <\mu} \mathrm{d}\omega\ g_ig_j.\\
\end{aligned}
\end{equation}

Similarly, the RKKY interaction between the same sublattice $B$, which is given by
\begin{equation}\label{SM_RKKY_BB}
\begin{aligned}
H^A_{\mathrm{RKKY}}&=-\frac{J^2}{\pi}\mathrm{Im} \int_{\epsilon <\mu} \mathrm{d}\epsilon\ \mathrm{Tr} [\bmS_2\cdot[\frac{\sigma_0-\sigma_z}{2}\otimes \bms] G_{0}^R(\RR_2,\RR_1,\epsilon )\bmS_1\cdot[\frac{\sigma_0-\sigma_z}{2}\otimes \bms] G_{0}^R(\RR_1,\RR_2,\epsilon )]\\
&=-\frac{J^2}{\pi}\mathrm{Im}\int_{\epsilon <\mu} \mathrm{d}\epsilon\ \mathrm{Tr} [\bmS_2\cdot  \bms G_{0,B}^R(\RR_2,\RR_1,\epsilon )\bmS_1\cdot\bms G_{0,B}^R(\RR_1,\RR_2,\epsilon )]
\end{aligned}
\end{equation}
where $G_{0,B}^R(\RR_2,\RR_1,\epsilon)$ is defined as
\begin{equation}
G_{0,B}^R(\RR_1,\RR_2,\epsilon)\equiv\langle \sigma_z=-1|G_{0}(\RR_1,\RR_2,\epsilon)|\sigma_z=-1\rangle=\ g_0 s_0 -g_1 s_2 -g_2 s_1.
\end{equation}
The RKKY interaction $H^B_{\mathrm{RKKY}}$ between the same sublattice B can also be divided into four type: the Heisenberg-type, the DM-type, the Ising-type and the anisotropic terms , of which strength can be calculated by
\begin{equation}\label{SM range function1}
\begin{aligned}
&\Lambda(\RR,\mu)=-\frac{J^2}{\pi}\mathrm{Im}\int_{\omega <\mu} \mathrm{d}\omega\ (g_0^2+g_1^2+g_2^2),\\
&\mathrm{D}_{1(2)}(\RR,\mu)=\frac{2 J^2}{\pi}\mathrm{Re}\int_{\omega <\mu} \mathrm{d}\omega\ g_0g_{2(1)},\\
&T_{ij}(\RR,\mu)=-\frac{J^2}{\pi}\mathrm{Im}\int_{\omega <\mu} \mathrm{d}\omega\ g_ig_j.\\
\end{aligned}
\end{equation}

Similarly, the RKKY interaction between the distinct sublattice $A$ and$B$, which is given by
\begin{equation}\label{SM_RKKY_BB}
\begin{aligned}
H^{AB}_{\mathrm{RKKY}}&=-\frac{J^2}{\pi}\mathrm{Im} \int_{\epsilon <\mu} \mathrm{d}\epsilon\ \mathrm{Tr} [\bmS_2\cdot[\frac{\sigma_0+\sigma_z}{2}\otimes \bms] G_{0}^R(\RR_2,\RR_1,\epsilon )\bmS_1\cdot[\frac{\sigma_0-\sigma_z}{2}\otimes \bms] G_{0}^R(\RR_1,\RR_2,\epsilon )]\\
&=-\frac{J^2}{\pi}\mathrm{Im}\int_{\epsilon <\mu} \mathrm{d}\epsilon\ \mathrm{Tr} [\bmS_2\cdot  \bms G_{0,AB}^R(\RR_2,\RR_1,\epsilon )\bmS_1\cdot\bms G_{0,AB}^R(\RR_1,\RR_2,\epsilon )]
\end{aligned}
\end{equation}
where $G_{0,AB}^R(\RR_2,\RR_1,\epsilon)$ is defined as
\begin{equation}
G_{0,AB}^R(\RR_1,\RR_2,\epsilon)\equiv\langle \sigma_z=+1|G_{0}(\RR_1,\RR_2,\epsilon)|\sigma_z=-1\rangle=\ g^\prime s_0.
\end{equation}
The RKKY interaction $H^{AB}_{\mathrm{RKKY}}$ between the distinct sublattice only contains the Heisenberg-type term, of which strength can be calculated by
\begin{equation}\label{SM range function1}
\begin{aligned}
\Lambda(\RR,\mu)&=-\frac{J^2}{\pi}\mathrm{Im}\int_{\omega <\mu} \mathrm{d}\omega\ g^{\prime2},\\
\end{aligned}
\end{equation}

\end{widetext}
\end{document}